\documentclass[preprint,aps,prb,endfloats]{revtex4}

\usepackage{epsfig}
\usepackage{amsmath}

\begin{document}

\title{Energy spectrum, persistent current and electron
localization in quantum rings}

\author{M. Manninen$^1$, P. Koskinen$^1$, M. Koskinen$^1$, P. Singha Deo$^2$, 
S.M. Reimann$^3$, and S. Viefers$^4$}

\address{$^1$NanoScience Center, Department of Physics, University of Jyv\"askyl\"a, Finland\\
$^2$S.N. Bose National Centre of Basic Science, Kolkata 98, India\\
$^3$ Department of Mathematical Physics, Lund University of Technology, Sweden \\
$^4$Department of Physics, University of Oslo, Norway\\}







\begin{abstract}
Energy spectra of quasi-one-dimensional quantum rings
with a few electrons are studied using several different
theoretical methods.
Discrete Hubbard models and continuum models are shown to give similar 
results governed by the special features of the one-dimensionality.
The energy spectrum of the many-body system can be described
with a rotation-vibration spectrum of a 'Wigner molecule' of 
'localized' electrons, combined with the spin-state determined from an
effective antiferromagnetic Heisenberg Hamiltonian.
The persistent current as a function of magnetic flux through the ring
shows periodic oscillations arising from the 'rigid rotation' of the 
electron ring. For polarized electrons the periodicity 
of the oscillations is always the flux quantum $\Phi_0$.
For nonpolarized electrons the periodicity depends on the 
strength of the effective Heisenberg coupling and changes
from $\Phi_0$ first to $\Phi_0/2$ and eventually to $\Phi_0/N$
when the ring gets narrower.
\end{abstract}

\maketitle

\section{Introduction}

Recent experimental developments in manufacturing
quantum dots\cite{tarucha1996} and rings\cite{lorke1998}
with only a few electrons have made 
quantum rings an ever increasing topic of experimental
and theoretical research. 
In a quantum ring the electrons move in a ring-shaped
quasi-one-dimensional confinement. The one-dimensionality
makes the electrons strongly correlated. Among the quantum effects seen
in such systems are the Aharonov-Bohm 
oscillations\cite{aharonov1959} and 
persistent currents\cite{buttiger1983}. 

Many properties of the quantum rings can be explained with 
single-electron theory, which in a strictly one-dimensional
(1D) system is naturally very simple. On the contrary,
the many-particle fermion problem in 1D systems is surprisingly
complicated due to enhanced importance of the Pauli exclusion
principle. 
It is then customary to say that strictly 1D systems are not  
'Fermi liquids' but 'Luttinger liquids' with 
specific collective excitations (for reviews 
see\cite{haldane1994,voit1994,schulz1995,kolomeisky1996}). 

\begin{figure}[ht]
\centerline{\epsfxsize=3.7in\epsfbox{mattif1.ps}}   
\caption{Two models of quantum  rings, a continuum ring and a ring
consisting of discrete lattice sites.}
\end{figure}

We consider two models for quantum rings.
In the {\it continuum model} the electrons move in an
external two-dimensional potential (shown schematically in Fig. 1a)
usually considered to be harmonic:
\begin{equation}
V(r)=\frac{1}{2}m_e\omega_0^2(r-R)^2,
\end{equation}
where $R$ is the radius of the ring and $\omega_0$ the strength
of the radial confinement. 
The electron-electron interaction is the normal long-range
Coulomb interaction.
If the number of electrons is small,
the many-electron states in this external
potential can be solved (numerically) exactly
using standard configuration interaction (CI)
methods\cite{chakraborty1995,niemela1996,koskinen2001}.

Another theoretical approach\cite{zvyagin1990,kusmartsev1991,yu1992}
 to quantum rings has
been a model
where the ring consists of
discrete lattice sites, as shown in Fig. 1b. 
The many-particle Hamiltonian can be approximated 
with the Hubbard model\cite{hubbard1963,peierls1933}
\begin{equation}
H = -t \sum_{i=1}^L \sum_{\sigma} 
  \left( e^{-i2\pi\phi/L} c_{i+1, \sigma}^{\dagger} c_{i, \sigma} 
  +  e^{i2\pi\phi/L} c_{i, \sigma}^{\dagger} c_{i+1, \sigma} \right)
 + U \sum_{i=1}^L \hat n_{i \uparrow} \hat n_{i\downarrow},
\end{equation}
where $t$ and $U$ are the Hubbard parameters determining the hopping 
between neighbouring sites and the on-site energy, $L$ is
the number of electrons and the number of sites, respectively, 
and $\phi$ is the magnetic flux through the ring (in units of 
the flux quantum $\Phi_0=h/e$).
The advantage of the discrete model
is that the many-body problem is
much easier than that of Eq. (1), and 
can be solved exactly in some limiting cases.

The two theoretical approaches, although seemingly very
different, give in many cases qualitatively similar results.
The aim of this work is to compare these models quantitatively
and study the reasons for the similarities of these
two approaches (for an introductory review see Ref.\cite{viefers2003}).

\section{Energy spectra: Rotational and vibrational states}

\begin{figure}[ht]
\centerline{\epsfxsize=3.7in\epsfbox{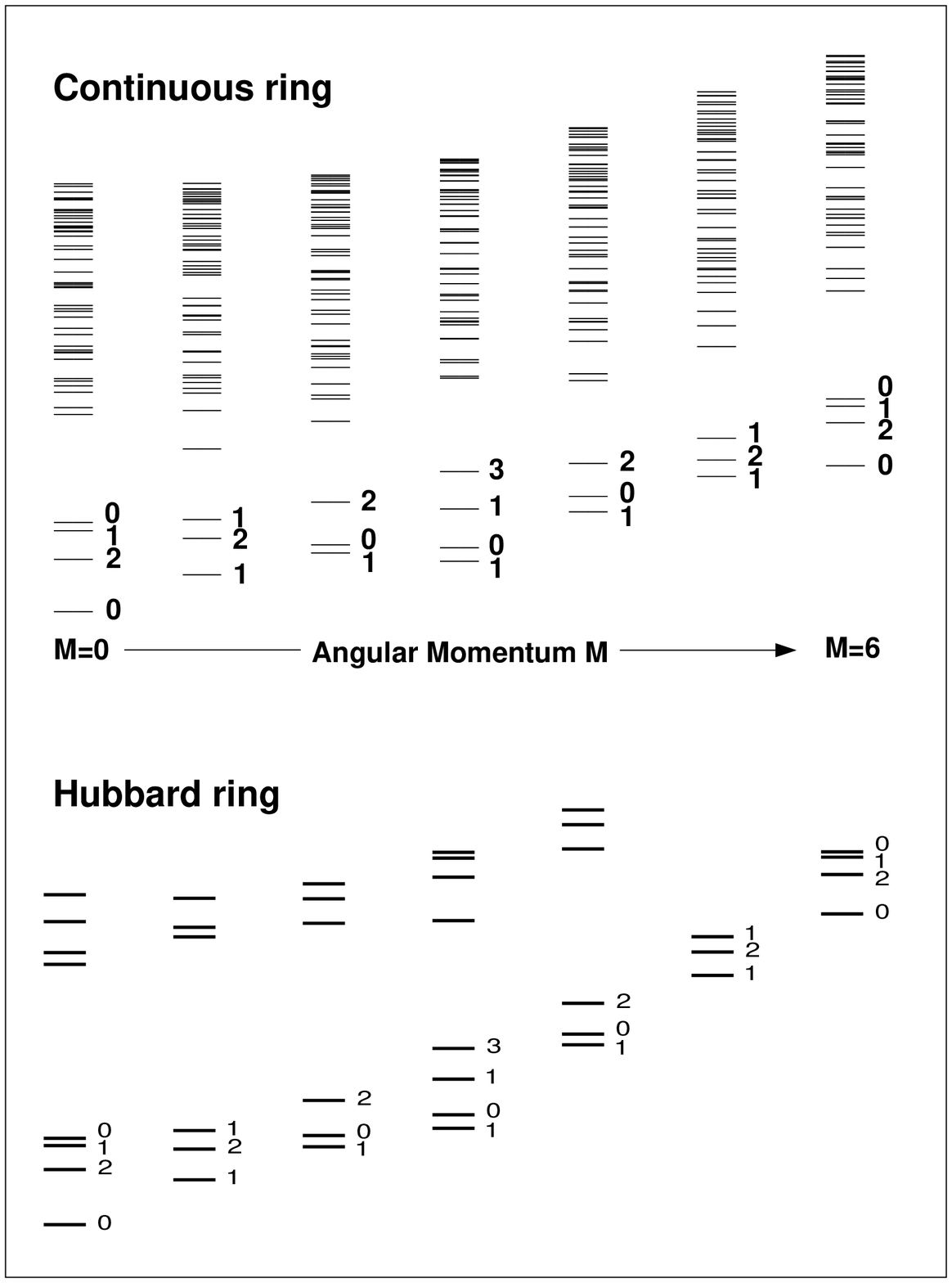}}   
\caption{Many-particle spectra of six electrons in a continuum ring and
in a Hubbard ring with eight sites. 
The numbers indicate the total spin of the state.}
\end{figure}

Figure 2 shows the energy spectra of quantum rings with six electrons,
calculated from the continuum model and from the lattice model
with eight sites. In both cases
the excitation spectrum contains a low energy band of
{\it rotational} states with energy increasing in the continuum model
roughly as $\hbar^2M^2/2I$, where $M$ is the total angular momentum and $I$
the moment of inertia $I=Nm_eR^2$. The low energy electron
spectrum thus corresponds to rigid rotation of a ring of six electrons
while the higher bands correspond to vibrational excitations.
Moreover, the energy splitting (due to spin) can be 
quantitatively described with an antiferromagnetic Heisenberg
model\cite{koskinen2001}.
Thus one arrives at the following model Hamiltonian
\begin{equation}
H_{\rm eff} = \frac{\hbar^2}{2I} {\bf M}^2 
	+ J\sum_{\langle i,j\rangle} {\bf S}_i\cdot{\bf S}_j
        + \sum_{\nu} \hbar \omega_{\nu} n_{\nu},
\end{equation}
where the last term describes vibrational states of electrons
localized on the ring (in Fig. 2 the vibrational states are the ones
not marked with the spin number).

Figure 2 shows that nearly exactly the same low energy spectrum can
be obtained from the Hubbard model with suitably chosen parameters.
It is known that in the limit of large $U$ the half-filled
Hubbard model ($N=L$) approaches the antiferromagnetic Heisenberg
model, explaining the correct spin structure of the model Hamiltonian
above. However, the correspondence seems to go even further:
If $L>N$ the rigid rotations and vibrational states also appear
in agreement with the continuum model. 

The similarity of the discrete lattice model with contact interaction
and the continuum model with long-range interaction can be traced
back to the special properties of one-dimensional systems.
The strong contact interaction effectively prevents electrons to pass
each other and the 'kinetic energy repulsion' makes the strong
$\delta$-function interaction look like a $1/r^2$ interaction
as evident from the Calogero-Sutherland model\cite{calogero1969,sutherland1971,viefers2003}.
Nevertheless, it is surprising that the similarity survives 
to quasi-one-dimensional rings considered in Fig. 2.

\begin{figure}[ht]
\centerline{\epsfxsize=3.7in\epsfbox{mattif3.ps}}   
\caption{Many-particle spectrum of four electrons in a ring
and a dot, shown in the insets ($U=40t$).}
\end{figure}

The 'localization' of electrons along the ring happens at all electron
numbers. Similar traces of electron localization can be found in the 
energy spectra of two-dimensional quantum dots\cite{manninen2001}.
Figure 3 shows the energy spectra of quantum rings
and dots with four electrons. In both cases the classical localization
geometry is a square. Indeed the low-energy spectrum is similar
for a ring and for a dot.
The Hubbard model gives again qualitatively the same spectra 
as the continuum model (not shown here)\cite{koskinen2001,viefers2003}.

\section{Periodicity of the persistent current}

The persistent current of a quantum ring can be determined from
the flux dependence of the total energy\cite{viefers2003}
\begin{equation} 
I(\Phi)= -\frac{\partial E}{\partial \Phi},
\end{equation}
where $\Phi$ is the magnetic flux through the ring.
Since the discrete Hubbard model gives the same energy levels as the 
continuum model, we can use it to study the persistent current.
The Hamiltonian (2) can be solved numerically for small number
of electrons and sites. Due to the phase factors the energy levels,
and consequently the persistent current, 
will be periodic functions of the flux. 

\begin{figure}[ht]
\centerline{\epsfxsize=3.7in\epsfbox{mattif4.ps}}   
\caption{Flux-dependence of the many-particle spectrum of a Hubbard ring
with eight sites and four electrons.}
\end{figure}

Figure 4 shows the spectrum of a Hubbard ring of four electrons 
in eight sites, as a function of $\Phi$ for different values of $U$.
For $U=0$ the ground state  energy has a periodicity $\Phi_0$.
When $U$ increases the period changes first to $\Phi_0/2$ and eventually
to $\Phi_0/N$. This happens at all electron numbers
and in a similar fashion for continuum\cite{koskinen2002} and 
discrete\cite{viefers2003} rings.
The increase of the strength of the radial confinement, $\omega_0$ in Eq. (1),
of the continuum model
corresponds to the increase of the on-site energy $U$ of the Hubbard model.
In both cases the ring becomes more strictly one-dimensional in the sense
that electrons are prevented to pass each other.

\subsection*{Acknowledgements}

This work has been supported  by the Academy of Finland under
the Finnish Centre of Excellence Programme 2000-2005 (Project No. 44875,
Nuclear and Condensed Matter Programme at JYFL).

\end{document}